\documentclass{iopjournal}
\usepackage[backend=biber,natbib,style=nature,doi=true]{biblatex}
\usepackage{amsmath}
\usepackage{siunitx}
\usepackage[version=4,arrows=pgf-filled,
mathfontname=mathsf]{mhchem}
\usepackage{subcaption}
\usepackage{xspace}

\newcommand{\me}{\ensuremath{m_\mathrm{e}}\xspace}
\usepackage{amsmath,amssymb}

\begin{document}

\articletype{Paper} 

\title{3D-printed components for electron–ion trapping: Pre-experimental tests of functionality and ultra-high vacuum compatibility}

\author{Vineet Kumar$^1$\orcid{0000-0001-8668-3663},
Niklas V. Lausti$^1$\orcid{0000-0001-9906-6971},
Jiří Hajnyš$^2$\orcid{0000-0002-9228-2521},
Ivan Hud\'{a}k$^{1,3}$\orcid{0009-0004-4162-6154},
David Moty\v{c}ka$^1$\orcid{0009-0005-9012-4721},
Adam Jel\'{i}nek$^1$\orcid{0009-0004-8659-4539},
Michal Hejduk$^{1,*}$\orcid{0000-0002-4417-4817}}

\affil{$^1$Charles University, Faculty of Mathematics and Physics, Dept.\ of Surface and Plasma Science, Prague 8, Czech Republic}

\affil{$^2$Faculty of Mechanical Engineering, VŠB - Technical University of Ostrava, Ostrava, Czech Republic}

\affil{$^3$Institute of Photonics and Electronics CAS, v.v.i., Chaberská 1014/57, Prague, Czech Republic}

\email{michal.hejduk@matfyz.cuni.cz}

\keywords{laser powder bed fusion, 3D printing, ion trap, ultra high vacuum}

\begin{abstract}
We demonstrate the ultra-high vacuum compatibility of a microwave-driven electron trap and an atomic oven (for atomic beam generation) fabricated through 3D printing via Laser Powder Bed Fusion (L-PBF). The trap integrates into a coaxial microwave cavity, enabling stable, narrow-band, high-amplitude oscillations of the electric field at the electrodes. The design also supports simultaneous trapping of ions. The oven performs well in ultra-high vacuum (UHV) environments without significant outgassing. In addition to achieving the UHV regime for 3D-printed components, pressure variations and their potential impact on electron-ion trapping experiments were investigated over a month. Our results show that experiments with electrons photodetached from trapped and laser-cooled ions are feasible with the trap and oven manufactured by the L-PBF method. These findings establish a foundation for future experiments in microwave detection and the study of low-energy ion-electron interactions at room temperature.
\end{abstract}

\section{\label{sec:intro}Introduction}

Electron traps are being explored for applications in detecting microwaves\cite{cridlandSingleMicrowavePhoton2016}, dark photons\cite{fanOneElectronQuantumCyclotron2022}, milli-charged dark matter particles\cite{carneyTrappedElectronsIons2021,budkerMillichargedDarkMatter2022}, and also in quantum computing\cite{haffner2008quantum}. A particular type -- the two frequency Paul trap\cite{foot_two-frequency_2018,mikhailovskii_trapping_2025} -- that concurrently traps ions is under investigation for creating antihydrogen atoms through the recombination of antiprotons and positrons\cite{dehmelt_economic_1995,leefer_investigation_2016}. The co-trapped system of ions and electrons may also facilitate the detection of radio-frequency and microwave signals through optical means, as described by \citet{lausti2025roadmap}. The setup presented here was developed primarily to test this hypothesis.

The two-frequency Paul trap combines electric fields for trapping both electrons and ions. Unlike the combined Paul-Penning trap by \citet{walz_combined_1995}, this method eliminates the need for magnets that cause Zeeman splitting of quantum energy levels of trapped ions. The associated reduction of the number of lasers necessary for laser cooling of ions is advantageous especially in experiments involving multiple optical transitions, such as quantum control of electron-ion recombination\cite{wolz_stimulated_2020,hudakMicrocavityIntegration2D2025}. Additionally, the risk of forming guiding centre atoms is also removed\cite{pohl_new_2006}.

Because electrons have lower mass than ions, microwaves are required to drive electron Paul traps. Our trap integrates into a coaxial microwave cavity\cite{jefferts_coaxial-resonator-driven_1995}, enabling stable, narrow-band, high-amplitude oscillations of the electric field at the electrodes. While traditional machining could produce this design\cite{osada_feasibility_2022}, the associated costs---particularly for prototyping and surface rounding to improve resonance quality---are prohibitive.

Additive manufacturing (AM) has significantly reduced expenses, with studies showing metal-coated 3D-printed plastic provides adequate performance metrics in some cases\cite{mohammed_3d_2021}. However, plastics cannot be used in ultra-high-vacuum (UHV) environments due to the requirement for baking above \qty{100}{\celsius}. Therefore, metal 3D printing, such as our selected method of Laser Powder Bed Fusion (L-PBF), is considered the only viable additive manufacturing approach.

This is especially true for atomic ovens. In experiments like ours\cite{matthiesen_trapping_2021,miossec_design_2022}, electrons and ions are produced by photo-ionization of neutral atoms (in our case, calcium) evaporated from a resistively heated dispenser. L-PBF was chosen to realize a design that achieves good thermal insulation while accommodating the tight space constraints of the vacuum chamber.

However, parts produced by L-PBF face potential issues. Porosity and unknown microvoids can outgas under UHV conditions, threatening the stability required for electron trapping. These voids might open upon heating, which is particularly problematic for our oven, as it operates above \qty{400}{\celsius}. Additionally, surface roughness from AM reduces the resonance quality factor of the coaxial resonator, necessitating inner surface polishing, which may not be realizable on a satisfactory level.

This study presents tests indicating that the identified concerns are unfounded. Vacuum compatibility assessments (Sections \ref{sec:UHV} and \ref{sec:UHV evolution}) of the experimental apparatus (described in Section \ref{sec:apparatus}) show that both the trap and oven manufactured by the L-PBF method (Section \ref{sec:componenets}) function effectively in UHV environments without notable outgassing or surface degradation, even at \qty{400}{\celsius} for the oven. Additionally, functionality tests of the trap as a microwave resonator (Section \ref{sec:Trch}) confirm the device's readiness for electron trapping experiments.

\section{Experimental apparatus}
\label{sec:apparatus}

\subsection{Design considerations}
\label{sec:design}

\begin{figure*}
    \centering
    \includegraphics[width=\linewidth]{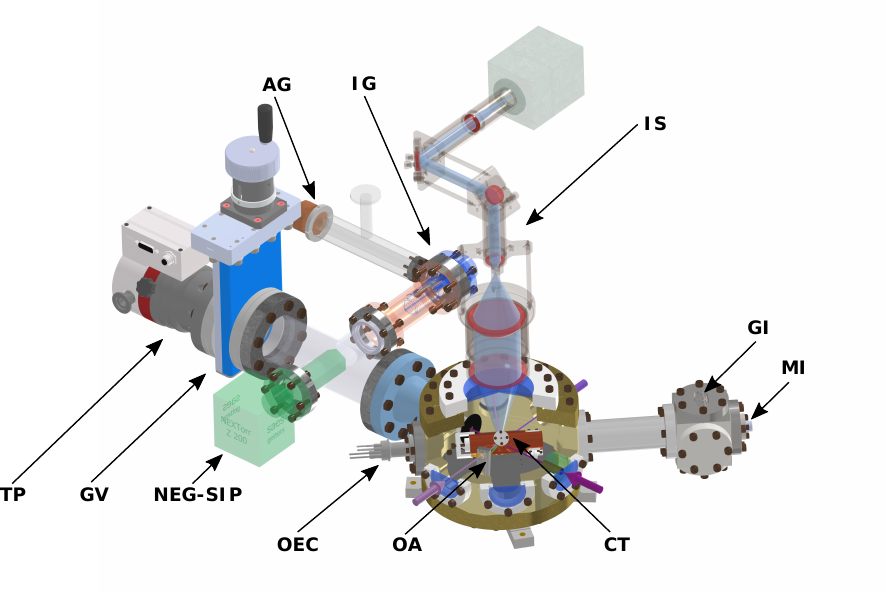}
    \caption{Schematic illustration of the EiTEx setup. The pumping system comprises a turbopump (TP) connected via a gate valve (GV) with an integrated sputter ion pump (SIP) and a non-evaporable getter (NEG). A cutaway view of the vacuum chamber highlights the oven assembly (OA) with electrical connections (OEC) and the coaxial trap (CT) equipped with megahertz (MI) and gigahertz (GI) signal inputs. A Pirani gauge (AG) and an ionization gauge (IG) are mounted on a side port of the vacuum chamber, while the imaging system (IS) is positioned vertically above the chamber.}
    \label{fig:apparatus}
\end{figure*}

Vacuum and electrical tests of 3D-printed parts are carried out using the Electron–ion Trapping Experiments (EiTEx) apparatus shown in Figure \ref{fig:apparatus}, illustrating the latest version that is ready for trapping experiments. Its design is dictated by the need to perform both optical and electrical measurements on electrons and ions confined in a dual-frequency Paul trap. Apart from achieving UHV conditions, the necessity of which is examined below, the apparatus also ensures  optical access to the trap interior enabling fluorescence imaging of laser cooled ions and operates without vibrations.

The requirement for UHV compatibility arises from the need to maintain full control over the trapped particle system. Specifically, the rate of momentum transfer collisions ($\nu_{\mathrm{MT}}$) between an electron and residual helium atoms must be significantly lower than the rate of any physically relevant processes ($\nu_i$) associated with the trapped electron quantum harmonic oscillator, i.e., $\nu_{\mathrm{MT}} \ll \nu_i$. If the device operates as a radio-frequency detector as described in \citet{lausti2025roadmap}, we require the ability to identify the spontaneous emission of the trapped-electron quantum harmonic oscillator, the rate of which is given by the formula\cite{fitzpatrick_quantum_2015}
\begin{equation}
    \nu_\mathrm{se} = \frac{n e^2 \omega_0^2}{6 \pi \epsilon_0 \me c^3}\,,
\end{equation}
where $n$ is the excitation level (principal quantum number), $e$ is the elementary charge, $\omega_0$ is the trapped electron's secular frequency, $\epsilon_0$ is the vacuum permittivity, and $c$ is the speed of light. For the bottom limit of our desired electron secular frequency $\omega_0\approx 2 \pi \times \qty{50}{MHz}$ and the excitation level on the order of 1000 quanta\cite{lausti2025roadmap}, the rate is on the order of \qty{E-4}{\per\second}. Electron state decoherence caused by other sources, such as black body radiation from the trap or electrical noise on trap electrodes, either have rates several orders of magnitude higher or can be suppressed through device design.

The enforcement of the condition $\nu_{\mathrm{MT}} \ll \nu_\mathrm{se}$ leads us to the upper pressure limit of $p_\mathrm{lim} \approx \qty{4E-8}{Pa}$ following the formula 
\begin{equation}
\label{eq:limpres}
    p_\mathrm{lim} = \frac{\nu_\mathrm{se}}{\sigma_{\mathrm{MT}}} \sqrt{\frac{\pi m k_B T}{8}}\,,
\end{equation} where $\sigma_{\mathrm{MT}}$ is the electron–helium momentum-transfer cross-section (with a value of \qty{5.2E-20}{\meter\squared} at room temperature $T = \qty{300}{\kelvin}$\cite{golden1966comparison}), $m$ is the helium atom mass, and $k_B$ is the Boltzmann constant. The equation can be derived from the standard kinetic theory of gases or following the reasoning by \citet{schut_expression_2025}.

In terms of electron confinement in Paul traps, experiments by \citet{matthiesen_trapping_2021} reported short lifetimes of electrons in the trap, approximately \SI{10}{ms}. This duration remained consistent despite variations in background pressure, ranging from \qtyrange{e-8}{e-6}{Pa}. As the authors suggest, this is likely attributable to alternate destabilization mechanisms, predominantly originating from the excitation of higher harmonics due to off-centre electron generation. In our experimental design, we aim to circumvent this issue by photo-detaching an electron from a laser-cooled ion situated at the centre of the trapping volume. Nevertheless, the newest experiments have achieved electron trapping durations that exceed \SI{2}{s} and indicate that surface contamination of the trap electrodes may be the limiting factor\cite{taniguchi_image_2025}. 

The primary source of surface contamination in ion trapping experiments is deposition of neutral precursor atoms. Therefore, their release has to be controlled. Laser ablation of a solid target provides excellent temporal control but requires direct optical access to the target and controllable focusing\cite{hendricks_all-optical_2007,duncan_invited_2012} to compensate for the limited spot lifetime\cite{battles_absorption_2024}. Additionally, it cannot guarantee ionic purity\cite{olmschenk_laser_2017,battles_absorption_2024} and scatters photons into fluorescence detection devices, necessitating automation due to increased background noise. In contrast, Joule heating of a powder dispenser is simpler and more cost-effective than the ablation method or laser heating of the dispenser\cite{gao_optically_2021} but struggles with controlling production time and amount. That can be achieved by a design of the oven that allows fast heating and cooling, and its rapid testing enabled by L-PBF manufacturing, which is the primary topic of this article.

To minimize the amount of light scattered toward sensitive photodetectors, it is necessary to incorporate beam blockers and beam dumps for both ionization and Doppler-cooling lasers. For the same reason, the thermal radiation from the oven must be also blocked as much as possible. Simultaneously, careful consideration must be given to the placement of high-reflection mirrors that form the optical cavity used in cavity quantum electrodynamics experiments with trapped particles\cite{takahashi_strong_2020}. These mirrors must be positioned to avoid contamination from particles emitted by the neutral atom source.

As a result, the system design is subject to multiple spatial constraints. Anticipating further miniaturization of the apparatus, we foresee the eventual need to adopt additive manufacturing techniques. This report therefore represents a foundational step in the ongoing development of our experimental platform.

\subsection{Components}
\label{sec:components}

The choice of components is made to meet the aforementioned requirements. The entire setup shown in Figure \ref{fig:apparatus} is horizontally mounted on a honeycomb optical table equipped with pneumatic vibration isolation (1HT, Standa Ltd.). The primary experimental components---namely, the coaxial particle trap, the oven assembly, and, prospectively, the in-vacuum optical cavity---are housed within a standard cylindrical chamber with a nominal diameter of \SI{160}{mm} and a height of \SI{85}{mm} (model number ZK200-8XCF63-01ALU, produced by VACOM GmbH), corresponding to an internal chamber volume of approximately \SI{2}{L}, with a total vacuum system volume of about \SI{3.96}{L}. To facilitate the achievement of ultra-high vacuum conditions, the experimental chamber is constructed of an aluminium alloy, which exhibits a lower \ce{H2} outgassing rate compared to stainless steel \cite{noauthor_outgassing_2016}. However, the maker recommends that the temperature of the alloy does not exceed \qty{120}{\celsius}, so the oven must be design to limit the thermal leakage and heat radiation to the chamber.

The need for optical access to the trap interiors is satisfied by the large borosilicate window (on a \qty{100}{mm} diameter Conflat flange, model number VPCF100B-L, produced by VACOM GmbH) on the top of the chamber. That, together with the size of the chamber, allows imaging of fluorescing ions using a lens with a numerical aperture of 0.375. Five of the eight available side windows are used for efficient short-range laser focusing.

The pumping section, connected to the aluminium chamber via a funnel-type reducer, is mainly composed of a standard stainless-steel (SS) DN63CF four-way nipple (420RKR063-040, Pfeiffer Vacuum). The pressure is monitored using a Pirani gauge (APG-1, Arun Microelectronics Ltd. -- AML) and a hot-filament ionization gauge (NGC3, AML). A turbomolecular pump (HiPACE 80 Neo, Pfeiffer Vacuum) with a pump speed of \SI{67}{\litre\per\second} for \ce{N2} establishes the high vacuum (HV) level. To achieve UHV, a combination of a non-evaporative getter pump (NEG) and a sputter ion pump (NEXTorr Z 200, SAES Getters S.p.A.) with Zr-V-Ti-Al getter alloy (ZAO) -- providing a pumping speed of \SI{290}{\litre\per\second} for \ce{H2} -- is employed. To prevent performance degradation from backflow, the turbomolecular pump is isolated from the apparatus by a manual UHV gate valve (GU0250MCCFM, Kurt J. Lesker) -- a bellows-sealed, copper-bonnet type with CF flanges made of 304 SS, featuring a linear-motion gate design. Once the valve is closed, the turbomolecular pumping stage is switched off so that it does not transmit vibrations.

\begin{figure}
    \centering
    \includegraphics[width=4in]{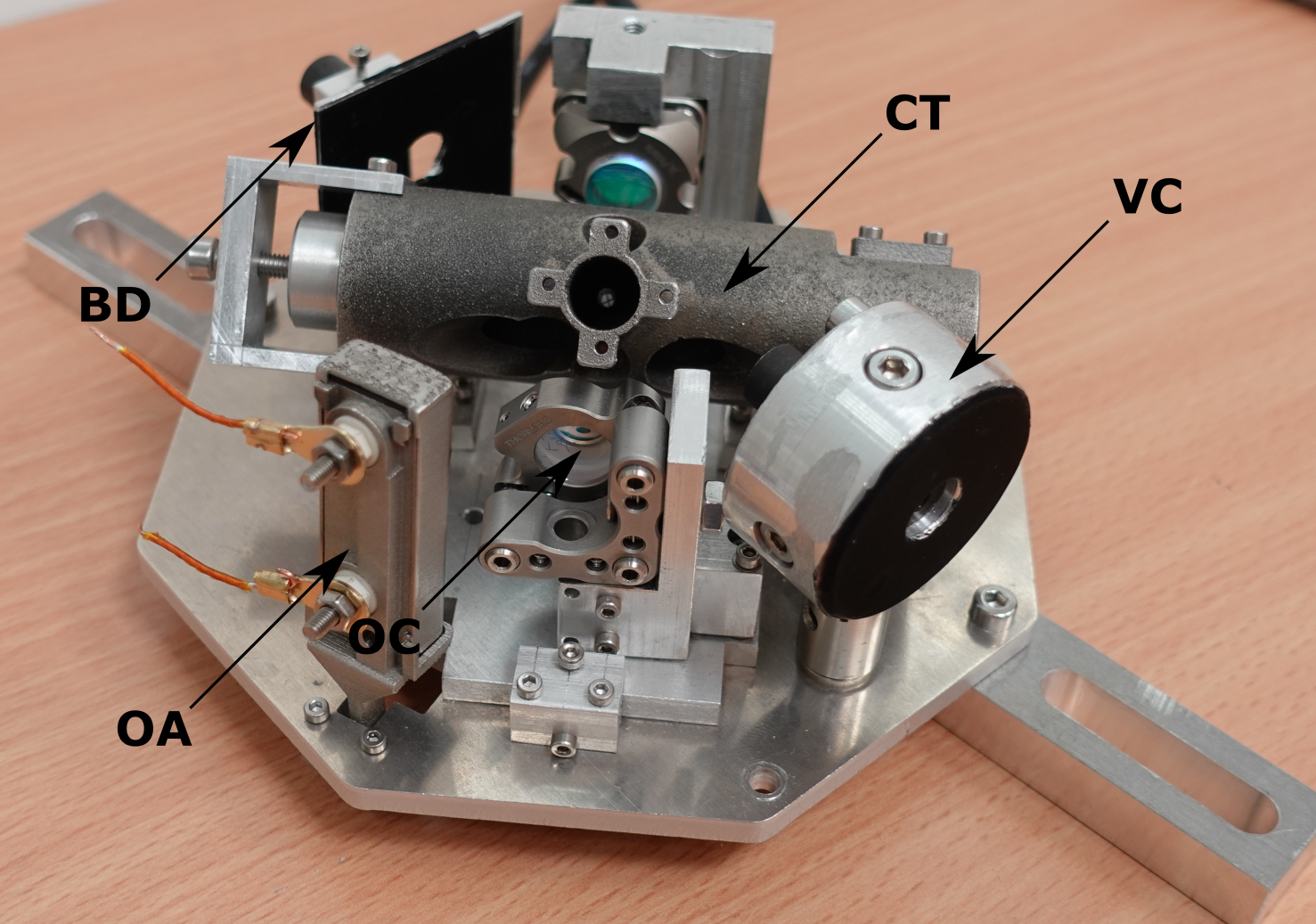}
    \caption{Photograph of the experimental setup showing the major subsystems mounted on a base plate (for public demonstration purposes). Key components are indicated: the oven assembly (OA) for atomic beam delivery, the coaxial trap (CT) for electron-ion trapping, the viewport coupler (VC) for optical access, the optical cavity (OC) for light–matter interaction, and the beam dump (BD) for blocking residual laser light}
    \label{fig:trapassembly}
\end{figure}

Figure \ref{fig:trapassembly} shows the full set of components present in the main vacuum chamber. The 3D-printed components (the trap and the oven assembly) are assembled on an aluminium base plate situated on the bottom blank flange of the cylindrical chamber. Following the design by \citet{kulkarni_ultrastable_2020}, the optical cavity is constructed from two kinematic mirror mounts (POLARIS-K05, Thorlabs) mounted on a custom-made posts incorporating a piezoelectric transducer (PU 40, piezosystem jena GmbH). The front sides of the viewport coupler (VC) and the beam dump (BD) are covered by Acktar Metal Velvet foil. Copper cylinders that constitute these parts are coated with cupric oxide dendrite that absorb diverging light rays\cite{norrgard2016vacuum}. 

\section{Materials and methods}
\label{sec:componenets}

In this section, we describe the design and manufacturing of the key components: the trap and oven assembly. The trap functions as a coaxial resonator, with the trapping region located at the antinode where the electric field is strongest. Unlike other comparable devices \cite{osada_feasibility_2022,mohammed_3d_2021}, our resonator features multiple openings in the cavity body to accommodate ionization and Doppler-cooling lasers, a neutral atom beam, an optical cavity, and fluorescence imaging. The openings for laser beams and fluorescence imaging have a conical shape, allowing for large acceptance and collection cones---an architecture that is readily achievable using the L-PBF method.

The oven assembly is positioned approximately \qty{5}{cm} from the trap centre and \qty{2}{cm} from the nearest chamber wall, making the suppression of heat leakage and thermal radiation essential. To address this challenge, the final design presented here was meticulously analyzed using finite-element simulations before being fabricated and tested. The key features that minimize thermal leakage are the heat shield and the narrow supporting leg. Additionally, to reduce production costs, we employed off-the-shelf ceramic electrical insulators, thereby avoiding expensive on-demand manufacturing of ceramic parts.

\subsection{Electron-ion trap}
\label{sec:electrontrap}

\begin{figure}
    \centering
    \includegraphics[width=6in]{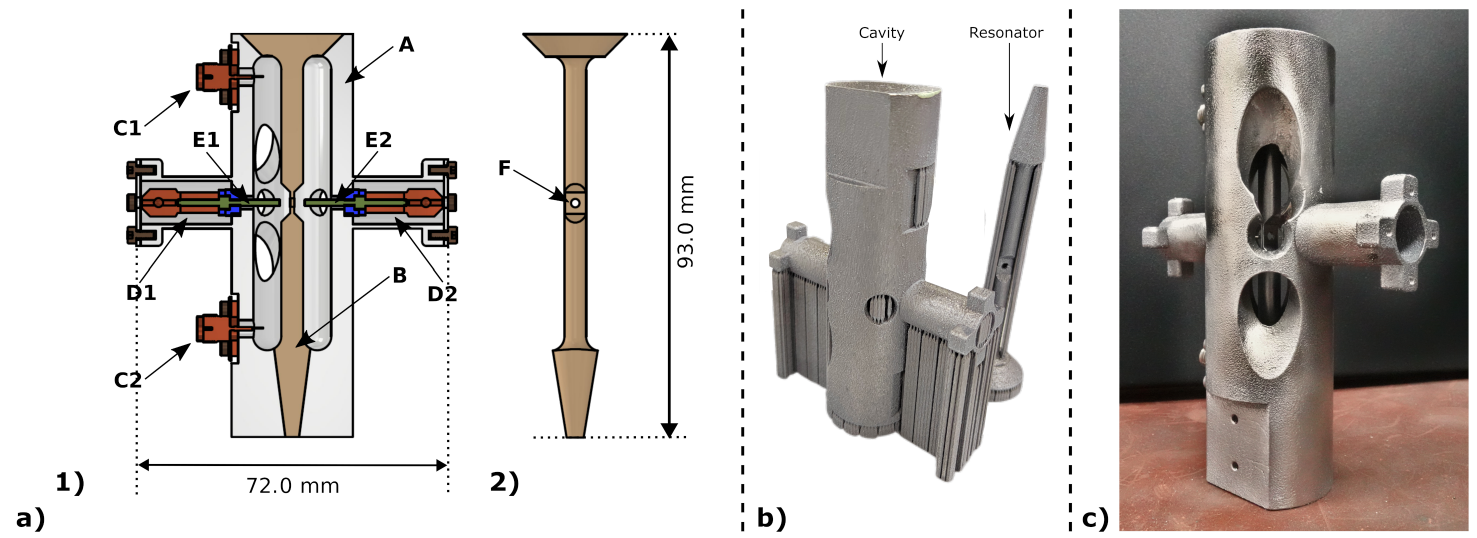}
    \caption{3D printed electron-ion trap designed for in-vacuum operation. (a) CAD model for printing and illustration purposes. (a1) Cross-sectional view of the outer body of cavity (A) with the resonator rod (B). Components (C1, C2) are SMA jacks for the GHz signal, used to capacitively drive the resonator, while components (D1, D2) are power push-on connectors for the MHz signal, connected to the end caps (E1, E2). (a2) Side view of the resonator, highlighting the trapping region hole (F). (b) Our 3D-printed assembly immediately after fabrication, with additive manufacturing support structures retained for mechanical stability during the printing process. (c) Trap after polishing and support removal, prepared for installation in the vacuum chamber.}
    \label{fig:trap}
\end{figure}

At the core of our experimental setup is a dual-frequency Paul trap illustrated in Figure \ref{fig:trap}. It is supplied with a high-frequency signal in the gigahertz range through an open coaxial connection (C1), with \SI{50}{\ohm} impedance, placed close to the node of the standing wave. The similarly situated coaxial jack C2 is used to monitor the resonance of the coaxial $\lambda/2$ resonator, which comprises the cavity body (A) and the resonator rod (B). Because of this resonance, a strong confining electric field is established within the aperture (F) located at the centre of the resonator rod. In addition, a low frequency signal in the megahertz range is applied to the end caps of the trap (E1 and E2) to facilitate the confinement of ions. The trap region (F) is defined by a cylindrical aperture with a radius of \SI{1.0}{mm}. The ratio of the inner radius of the resonant cavity (A) to the radius of the resonator rod (B) is $\SI{9.0}{mm} / \SI{2.5}{mm} = 3.6$, a value optimized by numerical simulations to maximize the quality factor of the resonator and, consequently, the amplitude of the trapping electric field.

The L-PBF technique was unsuitable for fabricating the depicted geometry as a single unit due to internal printing supports (Figure \ref{fig:trap}b) impeding removal within the cavity. Also, it would be difficult to polish the inner cavity surfaces with access only through laser beam apertures. Consequently, we divided the trap into two separate parts: the cavity body and the resonator rod. This approach preserves the resonator rod as a single entity, ensuring uninterrupted electric current flow and maintaining the quality factor\cite{mohammed_3d_2021}. To align the trapping region hole F axis with those of endcap electrodes E1 and E2, one interface between resonator rod B and cavity body A is crafted as a conical frustum with elliptical asymmetry, a feature unattainable by machining methods.

\sisetup{uncertainty-mode = separate}%

As the material for L-PBF, AlSi10Mg alloy powder with a grain size of approximately \qty{42\pm 22}{\micro\meter} was used, which inherently produced a relatively coarse surface finish. The cavity inner surfaces were therefore manually polished using FEPA\cite{StandardsFEPAFederation} grade P150 sandpaper and polishing cloth wheels (grit 120 and 180) until glaze was achieved (Figure \ref{fig:trap}c).

\begin{figure}
    \centering
    \includegraphics[width=5in]{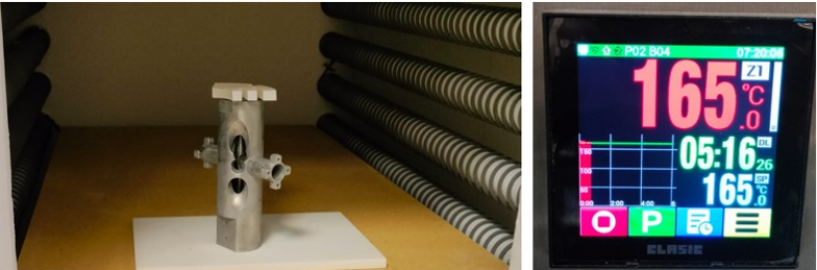}
    \caption{Baking of the EiTEx 3D-printed trap in a controlled oven. Left: Trap on a ceramic platform inside spiral heating coils for uniform heating. Right: Oven held at 165 $^\circ\text{C}$ for 5 h 16 m to ensure outgassing, structural stabilization and enhancement of electrical conductivity before vacuum integration.}
    \label{fig:Baking}
\end{figure}

Before installation into the vacuum chamber, the trap was first annealed in air at \qty{530}{\celsius} for \qty{30}{min}, followed by a prolonged annealing step at \qty{165}{\celsius} (Figure \ref{fig:Baking}). We aimed to enhance the material's electrical conductivity in accordance with the powder manufacturer's guidelines in this way, yet the resonator's quality factor remained unchanged in the later tests. After annealing, the trap was subjected to UHV-grade cleaning, which included sonication in deionized water at a frequency of \qty{37}{kHz} for 1 h at \qty{80}{\celsius}. This was followed by cleaning with high-pressure nitrogen gas and a final rinse in isopropanol.

\subsection{Atomic oven assembly}
\label{sec:oven}

\begin{figure}
    \centering
    \includegraphics[width=5in]{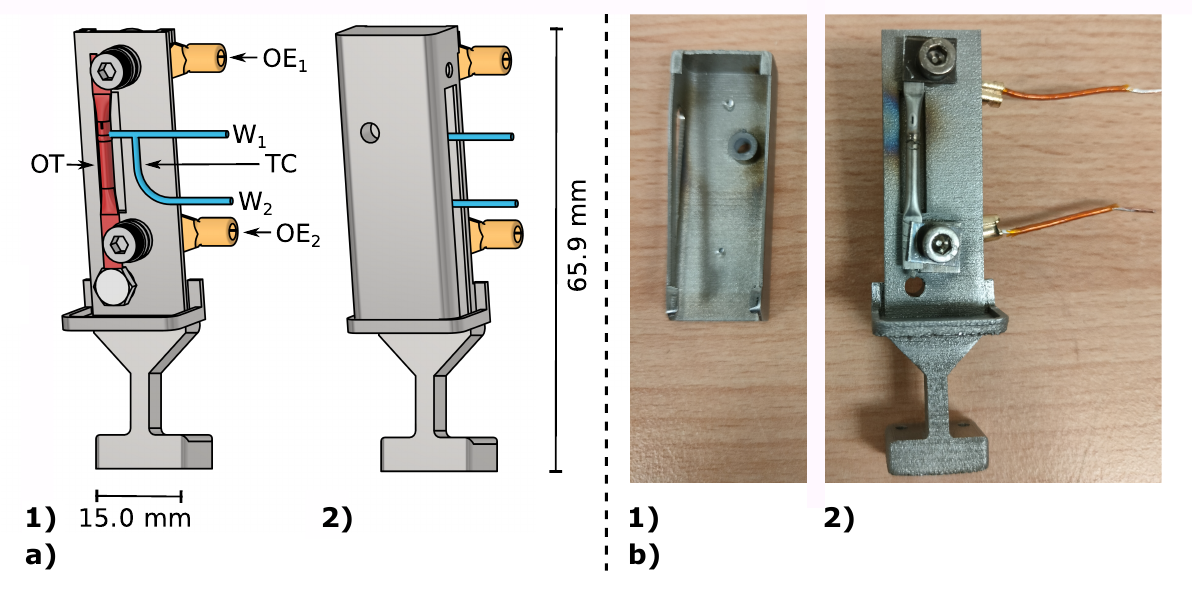}
    \caption{3D printed atomic oven assembly designed for in-vacuum operation. (a) CAD model for printing and illustration purposes. (a1) Front view of the assembly with the oven tube (OT) installed, displayed without the heat shield. $\text{W}_1$ and $\text{W}_2$ represent the thermocouple (TC) circuit wires, which are spot-welded to the oven tube for temperature monitoring. (a2) The assembly with the heat shield slid down from the top, fully enclosing the oven tube. $\text{OE}_1$ and $\text{OE}_2$ denote the electrical connections supplying power to the oven tube. (b) Photographs of the manufactured component: (b1) inner side of the heat shield, and (b2) assembled oven with electrical connections after operation in the vacuum chamber, where the darkened region on the surface is attributed to calcium evaporation and subsequent deposition during oven use, effectively preventing atomic contamination of the trap and other components inside the chamber.}
    \label{fig:cartridge}
\end{figure}

Figure \ref{fig:cartridge} illustrates the final design of the oven assembly. The oven tube (OT) is mounted on ceramic split bushes using stainless-steel screws connected to electrical contacts $\text{OE}_{1,2}$. For diagnostic purposes, a thermocouple (TC) is spot-welded to the OT in a configuration designed to minimize the influence of the heating current on temperature readings \cite{ballance_short_2018}. When an electric current is applied to the OT, the indium plug is initially melted, thereby opening a pathway for calcium vapours that are generated by continuing heating of the tube. To mitigate the risk that the melted indium might form a coating on the calcium inside the OT, the holder is inclined forward so that the indium is expelled from the tube. The thin supporting leg of the oven assembly serves to minimize thermal leakage and hence to achieve rapid heating of the OT. The L-PBF technique also allowed us to reduce the volume of material that is inadvertently heated.

Considering the criteria of a high melting point and low thermal conductivity, we selected 316L grade steel. The printing powder, possessing a grain size of \qtyrange{15}{45}{\micro\meter}, formed uniform, dense layers that closely corresponded with the observations by \citet{simchi2025mastering}, resulting in material with negligible voids.

After printing, the oven underwent sonication in deionized water at a frequency of \qty{37}{kHz} for 1 h at \qty{80}{\celsius}. This was followed by cleaning with high-pressure nitrogen gas and a final rinse in isopropanol.

\section{Results and discussions}
\label{sec:reanddi}
This section evaluates the UHV compatibility of 3D-printed components and details the oven assembly characterization. Subsequently, we examine the resonance behavior of the $\lambda$/2 resonator. All experiments were carried out without the optical cavity, so that we can study the influence of 3D-printed parts. Acktar films were left inside the chamber because we have not observed their influence on the pressure we measure.

\subsection{UHV attainment testing}
\label{sec:UHV}

\begin{figure*}
    \centering\includegraphics[width=4.75in]{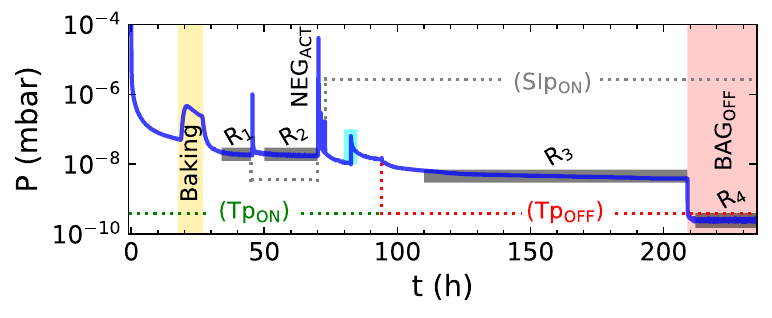}
    \caption{Pressure vs. time plot illustrating the progression of vacuum stages to achieve UHV conditions suitable for electron trapping. Initially, while $\mathrm{Tp}$ (green dotted line), pressure rapidly reduces to $5.0\times10^{-8}$  mbar. This is followed by the baking phase (yellow region), which significantly enhances outgassing. Subsequently, $\mathrm{SIp}$ (gray dotted line) is switched on, with a switch off during the NEG activation (indicated by a sharp peak due to desorption of gases released from the NEG material during its heating). The cyan-shaded region represents a phase of unknown outgassing, potentially caused by material-trapped gases or contamination from particles such as fingerprints, oil, or dust.  The red dotted line represents when there is no $\mathrm{Tp}$. Around $t \approx 200$ h, the BAG is switched off (pink region), and the final pressure drops to $<2.5\times10^{-10}$  mbar. The gray regions $(R_k)$ denote phases of nearly saturated pressure.}
    \label{fig:Pressuretime}
\end{figure*}

A multistep pumping process was carried out, involving pumping by a turbomolecular pump ($\mathrm{Tp}$) and a sputter-ion/NEG pump (marked as $\mathrm{SIp}$), and getter activation as shown in Figure \ref{fig:Pressuretime}. This process was carried out with the trap driving signals switched off and conducted under two configurations: with the gate valve $\mathrm{GV}$ open (corresponding to the $\mathrm{Tp_{ON}}$ phase in Figure~\ref{fig:Pressuretime}) and with the $\mathrm{GV}$ closed (the $\mathrm{Tp_{OFF}}$ phase). Initially, when $\mathrm{GV}$ was open and $\mathrm{Tp}$ was pumping, simultaneous pumping by sputter-ion pump was carried out for $\sim$ 24 h and then interrupted for 1h to activate the NEG. After $\sim$1.5 h, $\mathrm{SIp}$ was initiated, first $\sim$21.5 h with $\mathrm{Tp}$ and then with $\mathrm{GV}$ closed and no $\mathrm{Tp}$. At this point, the system relied solely on the $\mathrm{SIp}$, which continued to operate without introducing vibrational noise by the turbomolecular pump. This absence of vibrational noise is essential, particularly for the optical cavity sensitive to mechanical disturbances.

During the first $\sim\qty{100}{\hour}$ with the $\mathrm{SIp}$ active, the pressure remained stable. The Bayard-Alpert gauge (BAG) measured a pressure of $\sim\qty{4.0E-9}{mbar}$, while the sputter ion pump’s current gauge recorded a slightly lower pressure of $\sim\qty{1.9E-9}{mbar}$. Following the deactivation of the BAG, a notable pressure reduction exceeding one order of magnitude occurred, eventually stabilizing at \qty{2.5E-10}{mbar}, a value comparable to that used in SEM\cite{katagiri2006miniaturized} and LEED\cite{alakl2001electrochemical} experiments. The achievement of this pressure is noteworthy, considering the presence of various potential outgassing sources: besides the 3D-printed components (trap and oven), the radiation shield with cupric oxide dendrite, and the rough anti-reflection foil (Metal Velvet$^{\text{TM}}$ Black Foil, Acktar Ltd.) were installed in the chamber as well.

The duration of the process of $\sim$\qty{235}{h} can be significantly reduced by optimizing the ``holding'' intervals for the stabilization of pressure in four regions ($R_k; k=1,\ldots,4$), which currently accounts for more than $\sim$\qty{150}{h}. By implementing shorter intervals, the total time can be reduced to under \qty{50}{h}, enabling faster achievement of electron-trapping vacuum conditions.

\subsection{UHV evolution testing}
\label{sec:UHV evolution}

\begin{figure*}
    \centering
    \includegraphics[width=4.75in]{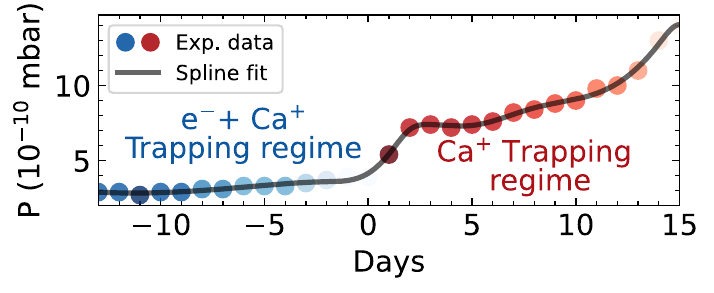}
    \caption{Long-term evolution of pressure in SIp regime. Prior to Day 0, the pressure remained below \qty{4E-10}{mbar}, a regime conducive to co-trapping electrons ($e^{-}$) and calcium ions ($\mathrm{Ca}^{+}$). Post-Day 0, the pressure increased by approximately an order of magnitude, transitioning to a regime that still permitted the trapping of $\mathrm{Ca}^{+}$.}
    \label{fig:Outgassing}
\end{figure*}

In addition to achieving the UHV regime for 3D-printed components, outgassing-induced pressure variations and their impact on the pressure requirements for electron–ion trapping were investigated over the course of a month. Throughout this time, the trap driving signals and turbomolecular pump remained off, while the UHV environment was sustained by the combination of the NEG and sputter-ion pumps. Figure \ref{fig:Outgassing} shows the long-term pressure evolution, demonstrating stable UHV conditions. This measurement was performed as a continuation of the previous test. 

The test began at a base pressure of $\sim 2.9\times10^{-10}$ mbar and, after nearly a month, rose to $< \qty{1.3E-9}{mbar}$, still demonstrating the UHV retention behaviour of our 3D-printed components. In this configuration, the sputter-ion/NEG pump remains the core pumping element. However, the finite sorption capacity of zirconium-based NEG alloys ultimately limits performance, as they absorb gases such as \ce{H2}, \ce{CO}, \ce{CO2}, \ce{O2}, and \ce{H2O}. During extended operation, species such as \ce{CO} and \ce{H2O} can saturate the bulk of the getter, reducing pumping efficiency and leading to a gradual pressure increase despite sputtering. This is probably the case of pressure increase after 2 weeks of operation in the SIp regime visible in pressure evolution graph Figure \ref{fig:Outgassing}.

This effect can be mitigated by periodic chamber baking with the turbomolecular pump on, which helps desorb accumulated gases. While large NEG pumps relative to chamber volume can maintain performance for long periods, smaller getters in gas-rich environments experience faster saturation and pressure drift.

During the deterioration of the vacuum level observed over a month, the system shifts from one experimental regime to another, a borderline between which is drawn by the line $p_\mathrm{lim} \approx \qty{4E-10}{mbar}$ from Equation \eqref{eq:limpres}. Below this limit, meaningful experiments with trapped-electron quantum harmonic oscillator in the presence of a co-trapped laser-cooled ion as described in \citet{lausti2025roadmap} can be carried out. Crossing this limit means shifting to a regime, in which only confinement of \ce{Ca+} ions makes sense to us. The upper limit for this pressure is on the order of  \qty{E-8}{mbar} as reported in experiments with \ce{Ba+} ions by \citet{sauter1986observation}.

\subsection{Oven testing}
\label{sec:Ovte}

\begin{figure}
    \centering
    \includegraphics[width=6in]{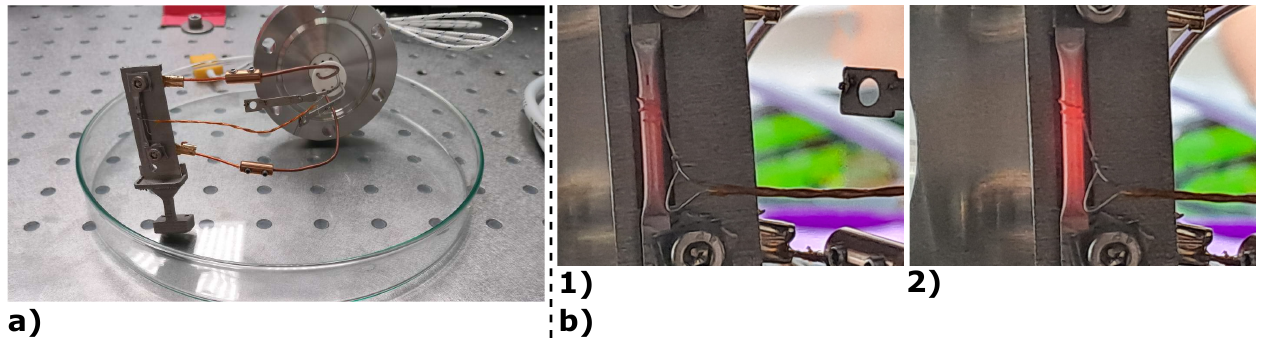}
    \caption{Atomic oven assembly installation and heating. (a) Oven assembly with electrical connections prior to installation, including a thermocouple for temperature measurement and power leads for tube heating. A heat shield is normally employed to reduce radiation losses but is omitted here for visual clarity. (b1, b2) Close-up views of the oven tube during operation, where the red glow under current supply indicates evaporation of atomic species. Of the total 32 mm heating length, an effective depot length of 11 mm was (geometry detailed in \cite{alfavakuo2024}) red glow in figure b1 shows nearly uniform heating below to the rectangular aperture at $\sim 460\, ^\circ\text{C}$. The transition from (b1) to the brighter glow in (b2) occurred within 2–3 s as the current increased by 1 A from 11 A, with the oven temperature in (b2) measured at $\sim 475\, ^\circ\text{C}$.}
    \label{fig:Oven glowing}
\end{figure}

Before installing the oven in the vacuum apparatus, we tested its heating and temperature monitoring capabilities in air, as shown in Figure \ref{fig:Oven glowing}a. The thermal radiation was confined to the oven tube (Figure \ref{fig:Oven glowing}b), and the attainment of temperatures higher than \qty{673}{\kelvin} indicated effective thermal insulation of the oven tube. This testing also verified the appropriateness of installing a heat shield.

The performance of an oven assembly under ultra-high vacuum conditions was examined by measuring dependence of the oven temperature ($T_\mathrm{ov}$) and the pressure on the input current ($I$), which was incremented in steps of $I_\mathrm{sz}$ held for durations of $t_\mathrm{hd}$. This stepwise heating prevented the pressure surges that could potentially collapse the vacuum system. During the tests, the system functioned with the gate valve open and the turbomolecular pump active, corresponding to the Tp$_\mathrm{ON}$ regime in Figure \ref{fig:Pressuretime}, with limit pressures approaching \qty{e-8}{mbar}.

\begin{figure*}
    \centering
    \includegraphics[width=5in]{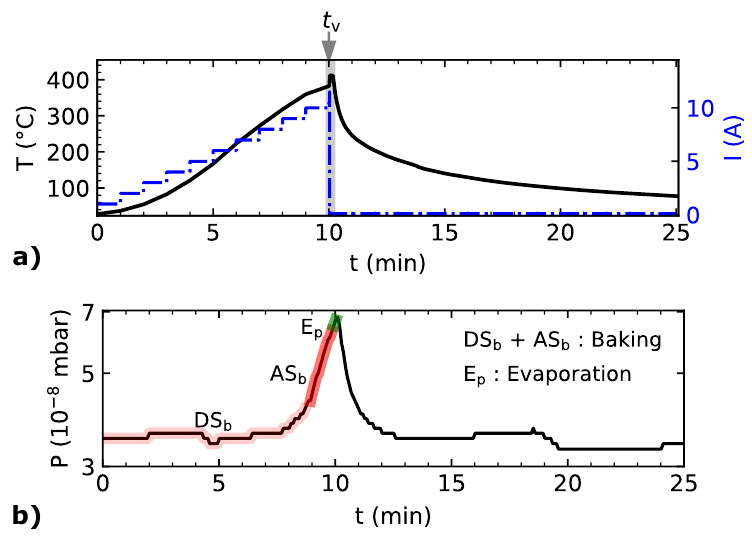}
    \caption{(a) Time evolution of oven temperature and applied current during a single calcium atomic beam pulse. During the initial $\sim$10 min baking phase, a stepwise current ramp steadily increases the oven temperature. At $t_{\mathrm{v}}=$10 min, a sharp 1 A increase in current drives the temperature above 410 °C, initiating rapid evaporation. The gray-shaded region serves as a visual guide for the evaporation pulse, occurring in phase with the current peak. Following this, the oven enters a cooling phase that follows a power-law decay behavior. (b) Corresponding pressure response measured in the vacuum chamber. During the ``dormant'' phase (DS$_\text{b}$), the pressure slowly rises \qtyrange[range-open-phrase = {\text{from }}]{3.5e-8}{4.3e-8}{mbar} over $\sim\qty{9}{\minute}$ as calcium begins to sublime. A short pre-evaporation interval (AS$_\text{b}$, $< \qty{1}{\minute}$) precedes the pressure peak associated with evaporation (E\(_{\mathrm{p}}\)), where the pressure spikes to $\sim\qty{6.8E-8}{mbar}$. This high-flux phase lasts for $\sim$5–6 s and is followed by a rapid pressure drop during the cooling phase. Together, these panels reveal the interplay between controlled heating, atomic release, and vacuum response dynamics.}
    \label{fig:fast oven heating}
\end{figure*}

The measurement recorded in Figure \ref{fig:fast oven heating} was performed with the calcium cartridge that had its indium seal disengaged beforehand. To reach target temperatures ($\sim \qty{400}{\celsius}$), current was increased in \qty{1}{\minute} long, \qty{1}{A} steps until reaching \qty{10.5}{A}, when we saw the temperature stabilizing at \qty{382}{\celsius}. Further current increase to \qty{11.5}{A} caused a rapid temperature spike, upon which we cut the current supply. The establishment of \qty{10.5}{A}-level pressure and temperature took less than \qty{5}{s}. The return of the pressure to the baseline took approximately \qty{3}{\minute} but reaching the thermal equilibrium required longer.

We identify the temperature spike period as the evaporation phase -- the point of rapid phase transition. The fact that this phase occupies only a fraction of the \qty{9}{\minute} long pressure variation cycle suggests optimization potential through controlled heating strategies, which remains a subject for further development.

\begin{figure*}
    \centering
    \includegraphics[width=5in]{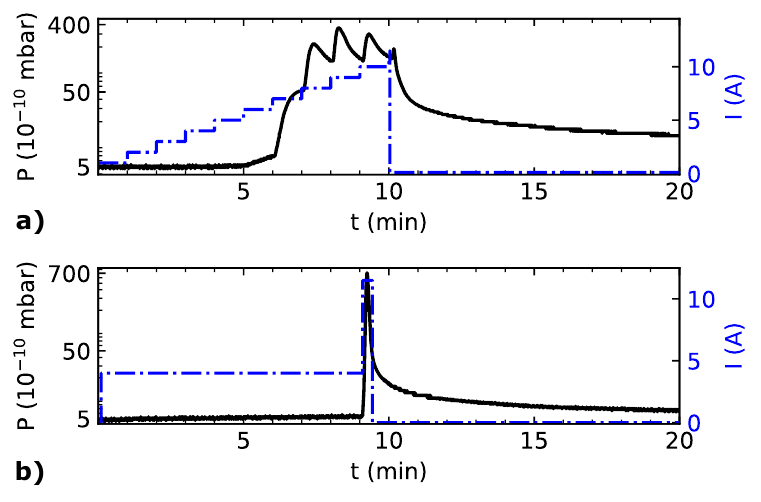}
    \caption{Temporal evolution of chamber pressure during two oven heating sequences under ultra-high-vacuum conditions. The heating process shown in the upper panel (a) corresponds to the sequence presented in Figure \ref{fig:fast oven heating}, showing an approximate 75-fold pressure increase. In the lower panel (b), the sequence exhibits a pressure rise of about 151-fold, followed by a rapid decrease to nearly 6-fold above the baseline within approximately 10 seconds after the peak, indicating a rapid burst-and-decay process completed within 30 seconds. The overall stability can be more effectively achieved by extending the oven tube baking time at a constant current and minimizing the evaporation phase.}
    \label{fig:Oven baking UHV regime}
\end{figure*}

Following the initial testing and characterization of the oven at higher pressures, current–pressure measurements were conducted under UHV conditions (Figure \ref{fig:Oven baking UHV regime}a). In a heating sequence of \qty{1}{\ampere\per\minute}, starting from a base pressure of \qty{4.9E-10}{mbar}, the pressure increased to \qty{3.6E-8}{mbar} before reaching \qty{10}{\ampere}, exhibiting three distinct peaks. The current was then increased to \qty{11.5}{\ampere} to confirm a pressure spike at this value, and was reduced to \qty{0}{\ampere} in \qty{5}{\second}. This behaviour indicates complex evaporation and pumping dynamics within the \qty{E-10}{mbar} pressure range achieved by the SIp and NEG in our setup. The turbomolecular pump, situated directly in line of sight from the trap centre with a higher limit pressure, naturally exhibits different pressure dynamics.

In another run starting from \qty{4.7E-10}{mbar} (Figure \ref{fig:Oven baking UHV regime}b), the current increased from \qty{0}{\ampere} to \qty{4}{\ampere} in \qty{6}{\second}, held for \qty{9}{\minute}, then increased to \qty{11.5}{\ampere} and reduced to \qty{0}{\ampere} in \qty{20}{\second}. During this process, the pressure rose to \qty{7.1E-8}{mbar} and dropped back to \qty{9E-10}{mbar} in about \qty{3}{\minute}. The narrower pressure spike during the fast scan from \qty{4}{\ampere} to \qty{11.5}{\ampere} and back to \qty{0}{\ampere} indicates a more confined, efficient evaporation phase with lower outgassing and faster vacuum recovery compared to the broader multi-peak response of the previous run.

The fact that thermal radiation is confined to the oven tube---while the surrounding components remain comparatively non-radiative---together with the rapid pressure and temperature spikes observed during heating at currents above \qty{4}{\ampere}, indicates that the pressure rise is unlikely to result from permeation of unknown molecules through the 3D-printed stainless steel. Instead, it is more plausibly attributed to evaporation of calcium powder. The three calcium evaporation tests helped determine the current threshold for evaporation and safe operation parameters. The measurements also illustrate that experience from other apparatuses is not directly transferable; the ability to evaporate atoms without exceeding pressures on the order of \qty{E-10}{mbar} depends on both the apparatus design and operational procedures \cite{matthiesen_trapping_2021}. In prospective experiments, short evaporation pulses are likely to be employed. However, a slow evaporation procedure characterized by a \qty{1}{\ampere\per\minute} current increase does not impede electron trapping experiments if electron detachment from laser-cooled ions is carried out once the vacuum system reaches its limit pressure.

\subsection{Trap characterization}
\label{sec:Trch}

To show that the trap's surface roughness does not degrade its performance, we measured the quality factor of the trap/coaxial resonator in the final setting in the vacuum chamber. The feeding antenna (C2 in Figure \ref{fig:trap}a) was connected by coaxial cables to the programmable signal generator (TGR2053, Aim-TTI Instruments) through a SMA vacuum feedthrough (242-SMAD50-C40-2, Allectra GmbH). The upper connector (C1) was connected to a power detector (ZX47-40-S+, Mini-Circuits) for signal analysis.

\begin{figure}
    \centering
    \includegraphics[width=3in]{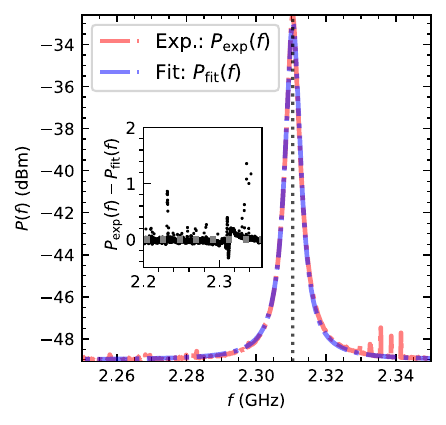}
    \caption{Measured power spectrum of the trap (red dashed), recorded with the trap installed in the vacuum chamber, as a function of frequency. The Lorentzian fit (blue dashed) determines the resonance frequency with a precision of approximately 4 kHz and achieves a determination coefficient \(R^2 > 99\%\), indicating excellent agreement with the measured data. The inset panel displays the residuals \(P_{\text{exp}} - P_{\text{fit}}\), showing only minor deviations in the spectral tails, consistent with statistical noise and confirming strong model fidelity. The minor peaks around 2.34 GHz are likely due to electromagnetic interference from the cellular network (n40 5G band). These peaks, at \qtyrange{-47}{-48}{dBm}, are \qtyrange{13}{14}{dB} weaker than the primary peak at \qty{-34}{dBm} and thus have a minimal impact on trap operation.}
    \label{fig:Resonancecurve}
\end{figure}

The power detector output is routed to a data acquisition card, allowing real-time monitoring and processing of the signal. A schematic of the experimental setup---used to characterize the resonator by applying a frequency-swept signal---is shown in \citet{laustifeasibility}. Resonance characterization is performed over a scanning range of \qtyrange{2.20}{2.36}{GHz}, with step size of \qty{100}{kHz}. Figure~\ref{fig:Resonancecurve} shows the resonance curve for electron trapping.

At an input power level of \qty{+13}{dBm}, the observed quality factor ($Q$) was $960$ at a resonant frequency of \SI{2.31}{GHz} when the device was physically attached to the chamber blank flange. In contrast, the resonator showed a $Q$ greater than $1050$ when not mounted inside the chamber, with resonance at \SI{2.345}{GHz}. These measurements show a reduction of a factor of two to three compared to those reported by \citet{mohammed_3d_2021} for a comparable 3D-printed apparatus and by \citet{taniguchi_image_2025} for an apparatus manufactured by conventional methods, which can be attributed to substantial apertures in the resonator cavity for optical accessibility. The disparity in values observed in and outside the vacuum chamber is likely due to signal reflections at the vacuum feedthrough interfaces. However, despite the reduced $Q$ factor, there is still significant amplification of the driving voltage signal, approximately by a factor of 30. This enables us to set the voltage amplitude on the trap-electrodes to approximately \qty{100}{\volt} (according to finite-element-method simulations)---an ideal value for electron trapping and resulting in the secular frequency $\omega_0$ larger than our minimum requirement of $2\pi\times \qty{50}{MHz}$ mentioned in Section \ref{sec:design}---while maintaining an input power of around \qty{1}{W} (\qty{+30}{dBm}), which is notably lower than used in experiments by \citet{matthiesen_trapping_2021}. 
 
To stress-test the microwave resonator, we applied a higher input power of \qty{10}{W} to the resonator situated in vacuum at the temperature of \qty{22}{\celsius}. Within three minutes of application, the pressure increased from \qty{6.1E-10}{mbar} to \qty{6.3E-10}{mbar}. Assuming a linear dependence of the measured pressure on the trap temperature---based on the hypothesis that particles with a thermal velocity distribution are released from the heated surface and the resulting pressure equilibrates with the surrounding gas---the pressure increase corresponds to a temperature rise of \qty{0.7}{\kelvin}.

The heating rate is double that observed in air with identical power input, which is anticipated due to the higher cooling efficiency in air. Therefore, we do not attribute the pressure increase to the release of molecules trapped in the 3D-printed material. Additionally, heating of coaxial cables by the microwaves cannot be reliably ruled out. Harmonic distortion of the input signal may also contribute to heating, as the resonator accepts only a portion of the input signal in this case.

Given the typical trapping times of less than \qty{1}{s} mentioned in Section \ref{sec:design}, we do not anticipate the need to operate the trap for periods exceeding \qty{1}{minute}. However, if this becomes an issue in later stages of the experiments, the choice of 3D-printed material may need to be revisited after considering all other aspects mentioned above.

\section{Conclusions and outlook}
\label{sec:conclusions}
In conclusion, we have introduced and characterized the EiTEx experimental setup, a system specifically developed for simultaneous trapping of an electron-ion system. The development relied heavily on 3D-printed components used in both the trap and oven assemblies, designed to meet stringent geometric and functional requirements. This advancement establishes a foundation for future experiments in microwave detection\cite{lausti2025roadmap} and the study of low-energy ion-electron interactions at room temperature\cite{mikhailovskii_trapping_2025}. Below, we present a concise summary of major findings and future perspectives.

\begin{enumerate}
    
    \item \label{item:first} The vacuum integrity of the system has been maintained at ultra-high vacuum conditions (UHV), reaching pressures as low as \qty{2.5e-10}{mbar}, with 3D-printed components (the trap and the atom oven). This was achieved using a vibrationless combination of a non-evaporable getter pump and a sputter ion pump and is comparable to other electron trapping apparatuses\cite{matthiesen_trapping_2021,mikhailovskii_trapping_2025}.
    
    \item \label{item:fourth} Generating an atomic beam---essential for ion trap loading---temporarily raises the vacuum chamber pressure beyond the threshold acceptable for quantum detection experiments with trapped electrons, as specified in Section \ref{sec:design}. Both the magnitude and duration of this pressure increase depend on the initial chamber pressure and the current ramping profile. Therefore, our experimental protocol must involve producing and storing sufficient precursor ions beforehand, re-establishing UHV conditions, and then conducting experiments with electrons. This approach is necessary because electrons formed under non-UHV conditions can be quickly lost. While there is potential to shorten the oven heating period, future detectors are likely to operate as described due to the time and energy required for atomic precursor production.
        
    \item \label{item:third} The coaxial resonator, integral to the Paul trap and constructed from AlSi10Mg alloy via 3D-printing, achieves a quality factor exceeding 960 at \qty{2.31}{GHz}. This is only up to 3.4 times lower than the quality factor reported by \citet{taniguchi_image_2025} for a closed, traditionally manufactured trap, despite its large optical access openings. Such a design facilitates electron trapping with minimal microwave input power.    
\end{enumerate}

These findings prove that experiments with electrons photodetached from trapped and laser-cooled ions are feasible with the trap and atomic oven manufactured by the L-PBF method.

\ack{We extend our gratitude to the Laborato\v{r} element\'{a}rn\'{i}ch proces\r{u} v plazmatu (EPP) group at the Faculty of Mathematics and Physics, Charles University, for their support. The Protolab 3D Printing Centre at the Technical University of Ostrava contributed the additive manufacturing of the components. This project also benefited from collaboration with the Cold Chemistry and AMO Physics group at the University of Liverpool.}

\funding{This work is supported by the Czech Science Foundation (GA\v CR: GA24-10992S), the Charles University Grant Agency (GAUK 295023 and GAUK 131224), and the Czech Ministry of Education, Youth, and Sports (project QM4ST, id. no. EH22\_008/0004572). Additional funding was provided by the Cooperatio Programme of Charles University. IH thanks to the
Technology Agency of the Czech Republic (TAČR: TN02000020) for the support. We also acknowledge previous funding from the University -- the Primus Research Programme (PRIMUS/21/SCI/005). }

\roles{Vineet Kumar: Conceptualization, Data curation, Formal analysis, Investigation, Visualization, Validation, Writing – original draft; Niklas V Lausti: Software, Validation, Visualization; Ji\v{r}\'{i} Hajny\v{s}: Resources; Ivan Hud\'{a}k: Data curation, Investigation; David Moty\v{c}ka: Investigation; Adam Jel\'{i}nek: Investigation; Michal Hejduk: Conceptualization, Funding acquisition, Project administration, Resources, Supervision, Writing – review \& editing}

\data{The data that support the findings of this study are openly
available\cite{vineetkumarSupportingDataTesting}.}


\printbibliography

@article{foot_two-frequency_2018,
  author = {Foot, C. J. and Trypogeorgos, D. and Bentine, E. and Gardner, A. and Keller, M.},
  journal = {Int. J. Mass Spectrom.},
  number = {430},
  pages = {117--125},
  title = {Two-frequency operation of a {Paul} trap to optimise confinement of two species of ions},
  volume = {430},
  year = {2018},
  doi = {10.1016/j.ijms.2018.05.007},
}

@article{leefer_investigation_2016,
  author = {Leefer, Nathan and Krimmel, Kai and Bertsche, William and Budker, Dmitry and Fajans, Joel and Folman, Ron and Häffner, Hartmut and Schmidt-Kaler, Ferdinand},
  journal = {Hyperfine Interact.},
  number = {1},
  pages = {12},
  title = {Investigation of two-frequency {Paul} traps for antihydrogen production},
  volume = {238},
  year = {2016},
  doi = {10.1007/s10751-016-1388-0},
}

@article{takahashi_strong_2020,
  author = {Takahashi, Hiroki and Kassa, Ezra and Christoforou, Costas and Keller, Matthias},
  journal = {Phys. Rev. Lett.},
  number = {1},
  pages = {013602},
  title = {Strong {Coupling} of a {Single} {Ion} to an {Optical} {Cavity}},
  volume = {124},
  year = {2020},
  doi = {10.1103/PhysRevLett.124.013602},
}

@article{matthiesen_trapping_2021,
  author = {Matthiesen, Clemens and Yu, Qian and Guo, Jinen and Alonso, Alberto M. and Häffner, Hartmut},
  journal = {Phys. Rev. X},
  number = {1},
  pages = {011019},
  title = {Trapping {Electrons} in a {Room}-{Temperature} {Microwave} {Paul} {Trap}},
  volume = {11},
  year = {2021},
  doi = {10.1103/PhysRevX.11.011019},
}

@article{taniguchi_image_2025,
  author = {Taniguchi, Kento and Noguchi, Atsushi},
  journal = {Physical Review A},
  number = {2},
  pages = {022420},
  title = {Image current detection of electrons in a room-temperature Paul trap},
  volume = {112},
  year = {2025},
  doi = {10.1103/6mpr-5s2h},
}

@article{battles_absorption_2024,
  author = {Battles, Kevin D. and McMahon, Brian J. and Sawyer, Brian C.},
  journal = {Appl. Phys. B},
  number = {12},
  pages = {214},
  title = {Absorption spectroscopy of {40Ca} atomic beams produced via pulsed laser ablation: a quantitative comparison of {Ca} and {CaTiO3} targets},
  volume = {130},
  year = {2024},
  doi = {10.1007/s00340-024-08332-8},
}

@article{miossec_design_2022,
  author = {Miossec, Chloé and Hejduk, Michal and Pandey, Rahul and Coughlan, Neville J. A. and Heazlewood, Brianna R.},
  journal = {Rev. Sci. Instrum.},
  number = {3},
  pages = {033201},
  title = {Design and characterization of a cryogenic linear {Paul} ion trap for ion–neutral reaction studies},
  volume = {93},
  year = {2022},
  doi = {10.1063/5.0080458},
}

@techreport{noauthor_outgassing_2016,
  institution = {VACOM Vakuum Komponenten \& Messtechnik GmbH},
  month = jun,
  number = {\#WP00002},
  type = {white paper},
  title = {Outgassing {Rates} of {Aluminum} compared to {Stainless} {Steel}},
  year = {2016},
  url = {https://www.vacom.net/fileadmin/user_upload/1.vacom.net/5.Downloads/5.5White_Paper/White_Paper_AluVaC_Ausgasverhalten_Web_EN.pdf},
}

@article{jefferts_coaxial-resonator-driven_1995,
  author = {Jefferts, S. R. and Monroe, C. and Bell, E. W. and Wineland, D. J.},
  journal = {Phys. Rev. A},
  number = {4},
  pages = {3112--3116},
  title = {Coaxial-resonator-driven rf ({Paul}) trap for strong confinement},
  volume = {51},
  year = {1995},
  doi = {10.1103/PhysRevA.51.3112},
}

@article{mohammed_3d_2021,
  author = {Mohammed, Ali Musa and Wang, Yi and Lancaster, Michael J.},
  journal = {Microw. Opt. Technol. Lett.},
  number = {3},
  pages = {805--810},
  title = {{3D} printed coaxial microwave resonator sensor for dielectric measurements of liquid},
  volume = {63},
  year = {2021},
  doi = {10.1002/mop.32679},
}

@article{ballance_short_2018,
  author = {Ballance, T. G. and Goodwin, J. F. and Nichol, B. and Stephenson, L. J. and Ballance, C. J. and Lucas, D. M.},
  journal = {Rev. Sci. Instrum.},
  number = {5},
  pages = {053102},
  title = {A short response time atomic source for trapped ion experiments},
  volume = {89},
  year = {2018},
  doi = {10.1063/1.5025713},
}

@article{norrgard2016vacuum,
  author = {Norrgard, Eric B and Sitaraman, Nathan and Barry, John F and McCarron, Daniel J and Steinecker, Matthew H and DeMille, David},
  journal = {Rev. Sci. Instrum.},
  number = {5},
  pages = {053119},
  title = {In-vacuum scattered light reduction with black cupric oxide surfaces for sensitive fluorescence detection},
  volume = {87},
  year = {2016},
  doi = {10.1063/1.4949503}
}

@misc{StandardsFEPAFederation,
  author = {FEPA},
  title = {Standards {\textbar} {{FEPA}} - {{Federation}} of {{European Producers}} of {{Abrasives}}},
  year = {2025},
  url = {https://fepa-abrasives.org/abrasives/standards/}
}

@article{katagiri2006miniaturized,
  author = {Katagiri, Souichi and Ohshima, Takashi},
  journal = {Microelectron. Eng.},
  number = {4-9},
  pages = {1058--1061},
  title = {Miniaturized electron gun for high-resolution scanning electron microscope using non-evaporable getter pumps},
  volume = {83},
  year = {2006},
  doi = {10.1016/j.mee.2006.04.004},
}

@article{alakl2001electrochemical,
  author = {Al-Akl, Ali and others},
  journal = {Phys. Chem. Chem. Phys.},
  number = {16},
  pages = {3261--3268},
  title = {Electrochemical and {UHV} characterisation of stepped {Pt} {100} electrode surfaces},
  volume = {3},
  year = {2001},
  doi = {10.1039/B100691K},
}

@inproceedings{laustifeasibility,
  author = {Lausti, Niklas and Kumar, Vineet and Hejduk, Michal},
  booktitle = {WDS'24 Proceedings of Contributed Papers — Physics},
  note = {33rd Annual Student Conference, Week of Doctoral Students},
  pages = {114--121},
  publisher = {MATFYZPRESS},
  title = {Feasibility of 3D-Printed Material for UHV and Microwave Electronics},
  year = {2024},
  isbn = {978-80-7378-520-8},
}

@article{haffner2008quantum,
  author = {H{\"a}ffner, Hartmut and Roos, Christian F and Blatt, Rainer},
  journal = {Phys. Rep.},
  number = {4},
  pages = {155--203},
  title = {Quantum computing with trapped ions},
  volume = {469},
  year = {2008},
  doi = {10.1016/j.physrep.2008.09.003}
}

@article{osada_feasibility_2022,
  author = {Osada, Alto and Taniguchi, Kento and Shigefuji, Masato and Noguchi, Atsushi},
  journal = {Phys. Rev. Res.},
  number = {3},
  pages = {033245},
  title = {Feasibility study on ground-state cooling and single-phonon readout of trapped electrons using hybrid quantum systems},
  volume = {4},
  year = {2022},
  doi = {10.1103/PhysRevResearch.4.033245},
}

@article{budkerMillichargedDarkMatter2022,
  author = {Budker, Dmitry and Graham, Peter W. and Ramani, Harikrishnan and {Schmidt-Kaler}, Ferdinand and Smorra, Christian and Ulmer, Stefan},
  journal = {PRX Quantum},
  number = {1},
  pages = {010330},
  title = {Millicharged {{Dark Matter Detection}} with {{Ion Traps}}},
  volume = {3},
  year = {2022},
  doi = {10.1103/PRXQuantum.3.010330},
}

@article{carneyTrappedElectronsIons2021,
  author = {Carney, Daniel and H{\"a}ffner, Hartmut and Moore, David C. and Taylor, Jacob M.},
  journal = {Phys. Rev. Lett.},
  number = {6},
  pages = {061804},
  title = {Trapped {{Electrons}} and {{Ions}} as {{Particle Detectors}}},
  volume = {127},
  year = {2021},
  doi = {10.1103/PhysRevLett.127.061804},
}

@article{fanOneElectronQuantumCyclotron2022,
  author = {Fan, Xing and Gabrielse, Gerald and Graham, Peter W and Harnik, Roni and Myers, Thomas G and Ramani, Harikrishnan and Sukra, Benedict AD and Wong, Samuel SY and Xiao, Yawen},
  journal = {Phys. Rev. Lett.},
  number = {26},
  pages = {261801},
  title = {One-electron quantum cyclotron as a milli-eV dark-photon detector},
  volume = {129},
  year = {2022},
  doi = {10.1103/PhysRevLett.129.261801}
}

@article{cridlandSingleMicrowavePhoton2016,
  author = {Cridland, April and Lacy, John Henry and Pinder, Jonathan and Verd{\'u}, Jos{\'e}},
  journal = {Photonics},
  number = {4},
  pages = {59},
  title = {Single {{Microwave Photon Detection}} with a {{Trapped Electron}}},
  volume = {3},
  year = {2016},
  doi = {10.3390/photonics3040059}
}

@inproceedings{hudakMicrocavityIntegration2D2025,
  title = {Microcavity Integration with {{2D Paul}} Trap},
  booktitle = {Quantum {{Optics}} and {{Photon Counting}} 2025},
  author = {Hud{\'a}k, Ivan and Kumar, Vineet and Lausti, Niklas and Honz{\'a}tko, Pavel and Hejduk, Michal},
  year = {2025},
  month = jun,
  volume = {13525},
  pages = {40--46},
  publisher = {SPIE},
  doi = {10.1117/12.3056538},
}

@article{dehmelt_economic_1995,
  author = {Dehmelt, Hans},
  journal = {Phys. Scr.},
  number = {T59},
  pages = {423},
  title = {Economic synthesis and precision spectroscopy of anti-molecular hydrogen ions in {Paul} trap},
  volume = {1995},
  year = {1995},
  doi = {10.1088/0031-8949/1995/T59/060},
}

@article{walz_combined_1995,
  author = {Walz, J. and Ross, S. B. and Zimmermann, C. and Ricci, L. and Prevedelli, M. and Hänsch, T. W.},
  journal = {Phys. Rev. Lett.},
  number = {18},
  pages = {3257--3260},
  title = {Combined {Trap} with the {Potential} for {Antihydrogen} {Production}},
  volume = {75},
  year = {1995},
  doi = {10.1103/PhysRevLett.75.3257},
}

@article{wolz_stimulated_2020,
  author = {Wolz, T. and Malbrunot, Chloé and Vieille-Grosjean, Mélissa and Comparat, Daniel},
  journal = {Phys. Rev. A},
  number = {4},
  pages = {043412},
  title = {Stimulated decay and formation of antihydrogen atoms},
  volume = {101},
  year = {2020},
  doi = {10.1103/PhysRevA.101.043412},
}

@article{pohl_new_2006,
  author = {Pohl, T. and Sadeghpour, H. R. and Gabrielse, G.},
  journal = {Phys. Rev. Lett.},
  number = {14},
  pages = {143401},
  title = {New {Interpretations} of {Measured} {Antihydrogen} {Velocities} and {Field} {Ionization} {Spectra}},
  volume = {97},
  year = {2006},
  doi = {10.1103/PhysRevLett.97.143401},
}

@article{simchi2025mastering,
  author = {Simchi, Abdolreza and Barthel, Bastian and Hein, Sebastian Boris and Reineke, Lea and Hosseini, Daniel},
  journal = {MRS Commun.},
  pages = {1--7},
  title = {Mastering binder jet 3D printing of 316L stainless steel: A descriptive model approach to particle size and deposition thickness},
  year = {2025},
  doi = {10.1557/s43579-025-00731-y},
}

@misc{alfavakuo2024,
  author = {{AlfaVakuo e.U.}},
  title = {Company Information},
  year = {2024},
  month = oct,
  day = {17},
  url = {https://alfavakuo.eu/?page_id=92},
  note = {Accessed on 3 May 2025}
}

@misc{mikhailovskii_trapping_2025,
  title = {Trapping of electrons and $^{40}$Ca$^{+}$ ions in a dual-frequency {Paul} trap},
  doi = {10.48550/arXiv.2508.16407},
  publisher = {arXiv},
  author = {Mikhailovskii, Vladimir and Sheth, Natalija and Qu, Guofeng and Hejduk, Michal and Lausti, Niklas Vilhelm and Satyajith, K. T. and Smorra, Christian and Werth, Günther and Yadav, Neha and Yu, Qian and Matthiesen, Clemens and Häffner, Hartmut and Schmidt-Kaler, Ferdinand and Bekker, Hendrik and Budker, Dmitry},
  month = aug,
  year = {2025},
  note = {arXiv:2508.16407 [physics]},
}

@article{sauter1986observation,
  author = {Sauter, Th and Neuhauser, Werner and Blatt, Rainer and Toschek, Peter E},
  journal = {Physical review letters},
  number = {14},
  pages = {1696},
  title = {Observation of quantum jumps},
  volume = {57},
  year = {1986},
  doi = {10.1103/PhysRevLett.57.1696},
}

@article{golden1966comparison,
  author={Golden, DE},
  journal={Physical Review},
  number={1},
  pages={48},
  title={Comparison of low-energy total and momentum-transfer scattering cross sections for electrons on helium and argon},
  volume={151},
  year={1966},
  doi={10.1103/PhysRev.151.48},
}

@misc{lausti2025roadmap,
  title = {Roadmap to planar electron-ion point Paul trap},
  doi = {10.48550/arXiv.2509.24396},
  publisher = {arXiv},
  author = {Lausti, Niklas Vilhelm and Kumar, Vineet and Hud{\'a}k, Ivan and Tarana, Michal and Hejduk, Michal},
  month = sep,
  year = {2025},
  note = {arXiv:2509.24396 [quant-ph]},
}

@book{fitzpatrick_quantum_2015,
  address = {Singapore},
  title = {Quantum mechanics},
  isbn = {978-981-4689-94-6},
  publisher = {World Scientific},
  author = {Fitzpatrick, Richard},
  year = {2015},
}

@article{hendricks_all-optical_2007,
  author = {Hendricks, R.J. and Grant, D.M. and Herskind, P.F. and Dantan, A. and Drewsen, M.},
  journal = {Applied Physics B},
  number = {4},
  pages = {507--513},
  title = {An all-optical ion-loading technique for scalable microtrap architectures},
  volume = {88},
  year = {2007},
  doi = {10.1007/s00340-007-2698-3},
}

@article{duncan_invited_2012,
  author = {Duncan, Michael A.},
  journal = {Review of Scientific Instruments},
  number = {4},
  pages = {041101},
  title = {Invited {Review} {Article}: {Laser} vaporization cluster sources},
  volume = {83},
  year = {2012},
  doi = {10.1063/1.3697599},
}

@article{olmschenk_laser_2017,
  author = {Olmschenk, S. and Becker, P.},
  journal = {Applied Physics B},
  number = {4},
  pages = {99},
  title = {Laser ablation production of {Ba}, {Ca}, {Dy}, {Er}, {La}, {Lu}, and {Yb} ions},
  volume = {123},
  year = {2017},
  doi = {10.1007/s00340-017-6683-1},
}

@article{gao_optically_2021,
  author = {Gao, S. and Hughes, W. J. and Lucas, D. M. and Ballance, T. G. and Goodwin, J. F.},
  journal = {Review of Scientific Instruments},
  number = {3},
  pages = {033205},
  title = {An optically heated atomic source for compact ion trap vacuum systems},
  volume = {92},
  year = {2021},
  doi = {10.1063/5.0038162},
}

@article{schut_expression_2025,
  author = {Schut, Martine and Andriolo, Patrick and Toroš, Marko and Bose, Sougato and Mazumdar, Anupam},
  journal = {Physical Review A},
  number = {4},
  pages = {042211},
  title = {Expression for the decoherence rate due to air-molecule scattering in spatial qubits},
  volume = {111},
  year = {2025},
  doi = {10.1103/PhysRevA.111.042211},
}

@article{kulkarni_ultrastable_2020,
  author = {Kulkarni, Soham and Umińska, Ada and Gleason, Joseph and Barke, Simon and Ferguson, Reid and Sanjuán, Jose and Fulda, Paul and Mueller, Guido},
  journal = {Applied Optics},
  number = {23},
  pages = {6999--7003},
  title = {Ultrastable optical components using adjustable commercial mirror mounts anchored in a {ULE} spacer},
  volume = {59},
  year = {2020},
  doi = {10.1364/AO.395831},
}

@misc{vineetkumarSupportingDataTesting,
  author = {Kumar, Vineet and Lausti, Niklas Vilhelm and Hud\'{a}k, Ivan and Moty\v{c}ka, David and Jel\'{i}nek, Adam and Hejduk, Michal},
  title = {Supporting {{Data}} for "{{3D}-printed components for electron–ion trapping: {Pre}-experimental tests of functionality and ultra-high vacuum compatibility}"},
  publisher = {National Repository Repo},
  year = {2025},
  doi = {10.48700/datst.74f00-v0b78},
  url = {https://data.narodni-repozitar.cz/general/datasets/74f00-v0b78}
}

\end{document}